\documentclass[aps,prb,showkeys,preprint]{revtex4}

\newcommand{\beq}{\begin{equation}}
\newcommand{\eeq}{\end{equation}}
\newcommand{\bef}{\begin{figure}}
\newcommand{\eef}{\end{figure}}

\def\bm{\boldmath}

\usepackage{graphicx}
\usepackage{amssymb}
\usepackage{pifont}
\usepackage{subfigure}

\begin{document}

\title{A superposition test for the emergence of nonlinearities \\ in a laser irradiated spherical absorber}
\author{Eshel Faraggi}
\email[To whom correspondence should be addressed: TEL: 317-587-0063, Electronic address: ]{efaraggi@gmail.com}
\affiliation{Research and Information Systems, LLC, Indianapolis, Indiana 46240}
\affiliation{Department of Physics, Indiana University Purdue University Indianapolis, Indianapolis, Indiana 46202}
\affiliation{Battelle Center for Mathematical Medicine, Nationwide Children's Hospital, Columbus, Ohio 43215}
\author{Bernard S. Gerstman}
\email{gerstman@fiu.edu}
\affiliation{Physics Department, Florida International University, Miami, Florida 33199}
\author{Andrzej Kloczkowski}
\email{andrzej.kloczkowski@nationwidechildrens.org}
\affiliation{Battelle Center for Mathematical Medicine, Nationwide Children's Hospital, Columbus, Ohio 43215}

\date{\today}

\begin{abstract}
The principal of linear superposition is investigated in the computational system of a solid spherical absorber immersed in a transparent aqueous medium and illuminated by a laser pulse. The absorber is exposed to a single top-hat pulse and to a fraction of the pulse from which a superimposed response is calculated. The results clearly show the transition of the system from a low fluence linear state where superposition is valid to a high fluence nonlinear state where the superposition is violated. The procedure described in the text can be used to find the transition to nonlinearity in a given excited system. This work can lead to better understanding and quantifying the transition from linear to non-linear regimes in complex systems. 
\end{abstract}

\maketitle

\vskip 1cm

{\bf We study the superposition principle for pressure waves in a computational model system of a spherical absorber immersed in water. This is a model for a laser irradiated melanosome found in the eye. We find a transition in the acoustic response between a linear state where the superposition principle holds to a nonlinear state where the superposition principle is violated. We estimate the critical fluence for this transition to be $10^{-3}$~J/cm$^2$.}

\vskip 2cm

The onset of nonlinear behavior in dynamical systems has intrigued scientist for many years.~\cite{suhl1956nonlinear,linsay1981period,testa1982evidence,feigenbaum1983universal,wolf1985determining,moses1986flow,hassanizadeh1987high,haan1989onset,kolodner1989dynamics,li1991extrinsic,pincus1991approximate,liu1993sound,epstein1998introduction,tang1999ultrasonic} For a review consult Schroeder.~\cite{schroeder2009fractals} The inherent complexity of the subject and the lack of analytical mathematical tools to analyze the dynamical equations result in  studies of simplified model systems  and reliance on numerical techniques. Typically a system would have a parameter or a set of parameters that variations of which will cause a transition between linear and nonlinear states. This transition is thought to be of vital importance in the  understanding and modeling of complex systems. Estimating the complexity of systems and identifying universal behavior is also a related problem.~\cite{feigenbaum1983universal,theiler1990estimating,uversky2016dancing}

In this paper the transition from a linear to a nonlinear state will be investigated for the commonly used system of a spherical absorber immersed in a liquid and illuminated by a pulsed laser.~\cite{comt96.1} This system is relevant for various applications, e.g., laser damage to the retina~\cite{cain2002thresholds}, material science~\cite{faraggi2007acoustical}, and other applications. It is also important because the  simplicity of this system makes it an excellent choice for investigations into the non-linear dynamics of fluids and complicated transitions to turbulence.
The system is composed of a spherically absorbing particle immersed in a transparent aqueous medium and illuminated by a pulsed laser.

The transition between a linear and nonlinear response for increasing absorbed energy will be shown. One of the most fundamental properties of linear systems is that of superposition. In formal terms the condition of linear superposition can be stated as follows. Let $R(x)$ be the response to stimulation $x$ and $R(y)$ be the response to stimulation $y$, then $R$ would be said to obey the principal of superposition if $R(x+y)=R(x)+R(y)$. One should note that linear systems obey the superposition principal exactly, while nonlinear systems generally do not. Hence the linear to nonlinear transition can be identified by observing where the principal of superposition is violated. For the specific system modeled here this transition will be studied as the   energy absorbed by the particle is increased, i.e., as the laser fluence is increased.

Linear systems are predictable. Predictability here will mean that a closed form expression can be written which will describe completely the state of the system, or that a sufficient number of measurements on the system will allow prediction of its future state. For linear systems this can be done, and for most reasonable linear systems closed form description of the state of the system can be found in practice. The next step beyond predictable is deterministic. A deterministic signal is taken here to be produced by a set of specific rules that uniquely determine the state of the system. A predictable signal is also deterministic, and can be either linear or nonlinear. However, a deterministic signal is not necessarily a predictable one, as in the case of a chaotic system for which no number of measurements allows prediction of its future state. One should note that in a given system the transitions from linearity to nonlinearity, from predictability to non-predictability, and the transition to chaos can all occur at different parameter values.~\cite{fara06.chaos}

The model~\cite{fara05.3} used in this work consists of a laser pulse incident on a uniform spherical absorber surrounded by a transparent medium.  The rate of energy absorption per unit mass is given by
\begin{widetext}
\beq
\label{erate}
\dot{I_e} = \frac{3 I_0}{4 a \tau_L \rho_0} [1- \frac{1}{2 \alpha_L^2 a^2} (1-e^{-2 \alpha_L a}(1+2 \alpha_L a))] = \frac{3 I_0}{4 a \tau_L \rho_0} C( \alpha_L a),
\eeq
\end{widetext}
where $I_0$ is the incident laser fluence in $Joule/cm^2$, $a$ is the radius of the absorbing sphere, $\tau_L$ is the laser pulse duration, $\rho_0$ is the static density of the sphere, and $\alpha_L$ is the absorption coefficient of the absorbing sphere.

The equations that govern the thermodynamic and mechanical behavior of the system are applied to each mass point inside the absorber and outside in the transparent medium. The mathematical dot operation $\dot{f}(t)$ means a total time derivative, and the spatial derivative with respect to the position {\bm $u$} of a specific mass point is denoted by $\nabla_{\mbox{\bm $u$}}$. The equation of motion is
\beq
\label{eom.femav}
\rho \ddot{\mbox{\bm $u$}} = - \nabla_{\mbox{\bm $u$}} P,
\eeq
where $P$ is the pressure and $\rho=\rho(t)$ is the time varying density. It is related to the static density by mass conservation
\beq
\label{mconv}
\rho_0 r^2 = u^2 \rho \frac{\partial u}{\partial r}
\eeq
where {\bm $r$} is the initial position of the mass element with $r$ its radial component and $u$ is the time varying radial component of the vector {\bm $u$}.

In finite-difference numerical implementation, changes in $\dot{u}$ are obtained using the relationship $\Delta \dot{u} = \ddot{u} \Delta t$, where the derivative is centered in the time interval. The updated values of $\dot{u}$ are then used in the equation of continuity to get the time rate of change of the specific volume, $v=1/\rho$
\beq
\label{eos.inp}
\dot{v}= v \nabla_{\mbox{\bm $u$}} \cdot \dot{\mbox{\bm $u$}},
\eeq
In a spherically symmetric system, this can be expressed as 
\beq
\dot{v}= \frac{v}{\rho_0 r^2} [2 \dot{u} u \frac{\partial u}{\partial r} + u^2 \frac{\partial \dot{u}}{\partial r}].
\eeq

Energy conservation in the absorber relates the rate of laser absorption to temperature rise, volume change, and heat loss through conduction to the surrounding medium. $\dot{v}$ from Eq.~(\ref{eos.inp}) is used with the conservation of energy to get the time rate of change of the temperature
\beq
\label{econs.in}
c_v \dot{T} = \dot{I_e} - B \alpha T \dot{v} + \frac{\lambda}{\rho} \nabla_u^2 T,
\eeq
where $c_v$ is the specific heat and $\lambda$ is the thermal conductivity of the absorber. It is assumed that the absorbing sphere has a constant bulk modulus $B$ and constant thermal expansion coefficient $\alpha$. The $\dot{v}$ from Eq.~(\ref{eos.inp}) and $\dot{T}$ from Eq.~(\ref{econs.in}) are then used to get the time rate of change of the pressure at locations inside the absorber with a linear equation of state (EOS):
\beq
\label{eos.in}
\frac{\dot{P}}{B}  = -\frac{\dot{v}}{v} + \alpha \dot{T}.
\eeq
The time evolution of the absorber is calculated using the updated values of $P$ in Eq.~(\ref{eom.femav}) to repeat the cycle for the next time step.

Outside in the medium, the EOS takes on a modified form.~\cite{haar84} It is assumed that the medium is softer and more compressible than the absorber and neither $B$ nor $\alpha$ remain constant. Conservation of energy in the transparent medium ($\dot{I_e} = 0$) is represented by
\beq
\label{econs.out}
T \dot{s} = \frac{\lambda}{\rho} \nabla_u^2 T,
\eeq
Updated values for the specific volume are obtained from increments of the form, $\Delta v = \dot{v} \Delta t$ where $\dot{v}$ is obtained from Eq.~(\ref{eos.inp}). The new values of $s$ and $v$ are then used in the EOS to obtain new values of $T$ and $P$. The time evolution of the medium is calculated using the updated values of $P$ in Eq.~(\ref{eom.femav}) and the updated values of $T$ in Eq.~(\ref{econs.out}) repeating the cycle for the next time step.

The specific details of the system studied are as follows. The absorber is chosen with parameters appropriate for melanosome, a major constituent of the retinal pigment epithelium of the eye. For the melanosome: radius $a=1 \; \mu$m, bulk modulus, $B=39.4$~GPa, density, $\rho=1.35$~g/cm$^3$, specific heat, $c_v = 2.51$~J/gK, thermal expansion coefficient, $\alpha = 2.98 \times 10^{-5}$~K$^{-1}$, and absorption coefficient, $\alpha_L = 10,000$~cm$^{-1}$. The thermal conductivity is taken as $\lambda = 5.56 \time 10^{-3}$~J/cmKs for both the melanosome and the aqueous medium. The initial density of the aqueous medium is taken as $\rho_0 = 1.0$~g/cm$^3$. The other parameters for the aqueous medium are incorporated into the EOS.~\cite{haar84}

Superposition is a property of linear systems. It is typically not obeyed by nonlinear systems. In this work it will be shown how the system described so far, looses its superposition property as the fluence is raised, indicating a transition to nonlinearity. In a system where superposition is valid, any process can be attributed to the linear sum of its subprocesses. For example, in the absorber system described in this text, superposition would require that we can view the response from a given pulse as a sum of any of its divisions.

In formal notation we first introduce $L[t_i,t_f]$ to stand for a tophat laser pulse with a given intensity that is turned on from an initial time $t_i$ to a final time $t_f$. We also define $\delta p_{L[t_i,t_f]}(t)$ as the pressure change generated by this laser pulse, for example at a position $u=2a$. Finally, for a given time interval $[0,\tau_L]$ over which the laser is turned on we introduce a sectioning of the interval, $t_1,t_2,t_3,\cdots , t_n$, such that $0=t_1 < t_2 < \cdots < t_n = \tau_L$. Then, if the system obeys the superposition principal we have
\beq
\label{supos.eq}
\delta p_{L[0,t_L]}(t) = \sum_{i=1}^{n-1} \delta p_{L[t_i,t_{i+1}]}(t).
\eeq

For vanishing fluence we of course expect a trivially linear response for the system presented, and with the idea of continuity we can expect a range of fluences where the system is linear to a good approximation. As we increase the fluence this linear approximation becomes less and less reliable, and as it will now be shown, the principal of superposition is increasingly more violated.

Out of the four thermodynamical variables used in this study, the pressure in the most responsive to laser excitation, undergoing the largest relative change. This is due to the high values for the bulk modulus for both the absorber and the medium. Hence, the pressure will be the variable of choice to be analyzed in this work. Deviation from superposition were more difficult to observe in quantities such as the temperature and the specific volume, again because of the large bulk modulus.

We begin in Fig.~\ref{sp.j5} with a comparison of the pressure response at $u=2~\mu$m caused by a single laser pulse of fluence $10^{-5}$~J/cm$^2$ and duration $\tau_L = 1$~ns labeled as single, and the superimposed signal from two example sectioning. The first with $t_i = 100 \cdot i$~ps labeled as 100ps. In practice instead of simulating the response from the nine different laser intervals, the response to the first laser section is calculated, i.e., in the interval $[0,100]$~ps, and time shift symmetry is used to calculate the other intervals. The response to the nine sections is then summed. The other sectioning given in Fig.~\ref{sp.j5} is for the same laser pulse but with $t_i = 1000 \cdot i / 3$~ps. It is labeled as 333ps in the figure. As before, for the superimposed lines we perform the sum on the right hand side of Eq.~(\ref{supos.eq}), using the time shift symmetry $\delta p_{L[t_i,t_{i+1}]}(t) = \delta p_{L[t_1,t_2]}(t-t_i)$ simulating only the initial $333$~ps part of the full 1~ns pulse. 
\bef[th!]
\includegraphics[width=1\textwidth]{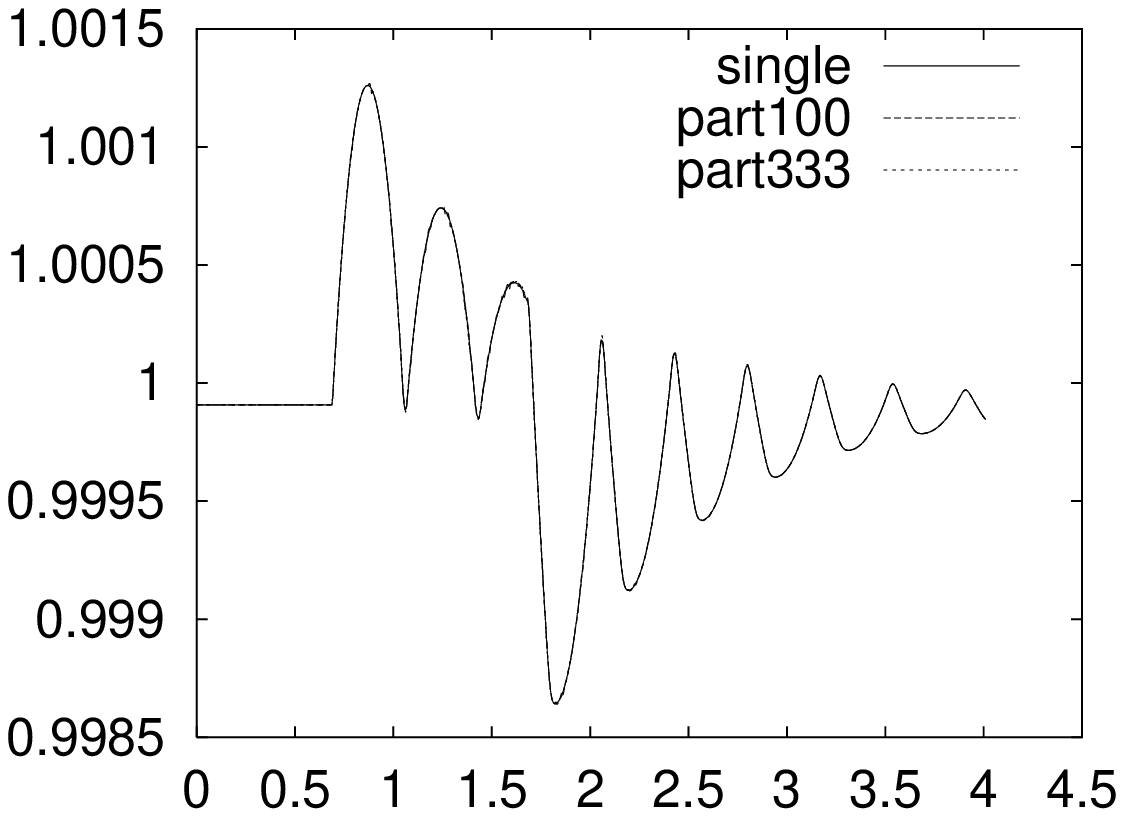} \\
\caption{Pressure at a location $u=2a$ as a function of time obtained from direct simulation and by applying the superposition principle with 10 sections of the original 1~ns laser pulse. The fluence of the laser pulse is $I_0 = 10^{-5}$~J/cm$^2$, for this fluence the two curves are identical and the  superposition principal is obeyed.}
\label{sp.j5}
\eef

As is evident from the plot the three lines overlap exactly and the superimposed responses give a faithful representation of the original pulse. This means that for the low fluence used here, $I_0 = 10^{-5}$~J/cm$^2$, the system behaves linearly and superposition in the pressure response is obeyed. One could use this type of behavior to calculate long term behavior of a system from only partial information regarding its initial behavior.

In Fig.~\ref{sp.j3} we repeat the process but this time with a higher fluence of  $10^{-3}$~J/cm$^2$. The duration of the pulse and of the sectioning are kept the same. For this fluence case we see that superposition is beginning to be violated, i.e., nonlinearities are beginning to creep up in the system. However, even for this case these deviations from nonlinearity are still small. 
\bef[th!]
\includegraphics[width=1\textwidth]{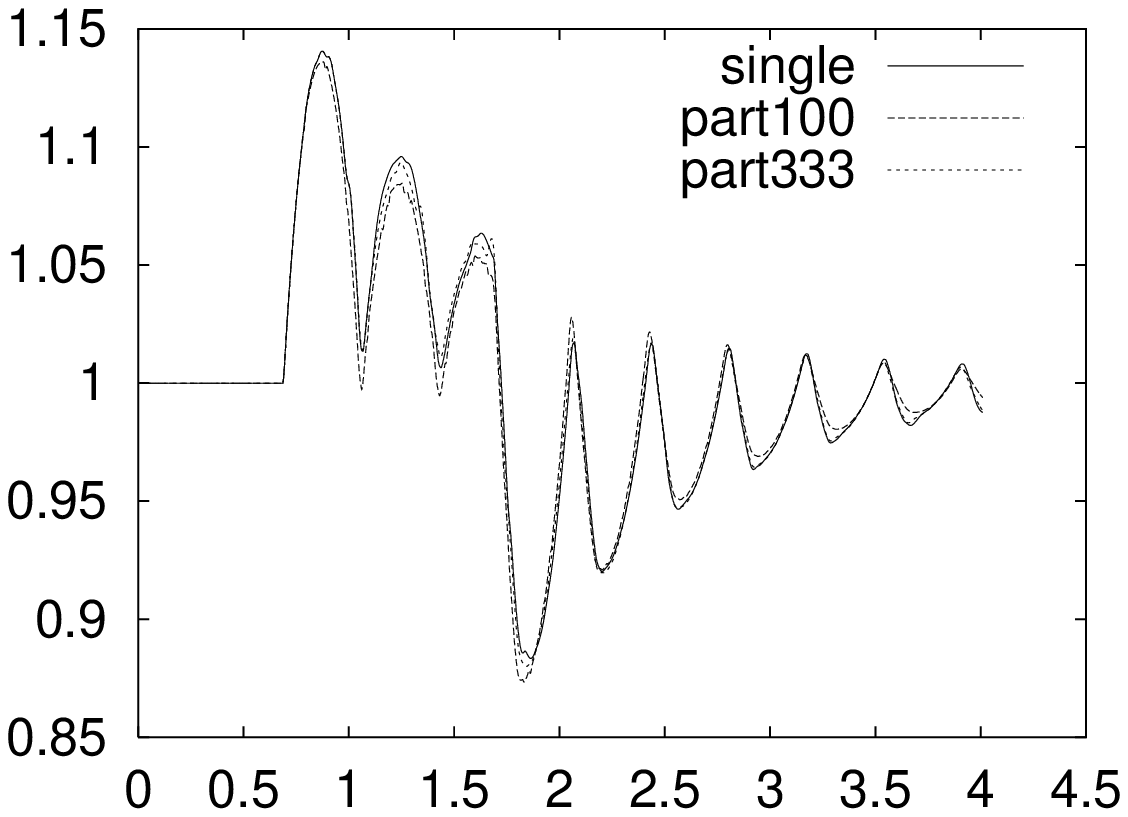} \\
\caption{Pressure at a location $u=2a$ as a function of time obtained from direct simulation and by applying the superposition principle with 10 sections of the original 1~ns laser pulse. The fluence of the laser pulse is $I_0 = 10^{-3}$~J/cm$^2$, for this fluence deviations between the two curves are starting to become evident though the  superposition principal is approximately valid.}
\label{sp.j3}
\eef

In Fig.~\ref{sp.j1}, we do the same procedure but this time with a fluence of 0.1~J/cm$^2$. For this higher fluence it is observed that the system is nonlinear and that superposition is no longer obeyed. Note that the initial delay is the figures results from the time is takes the acoustic signal to reach the point at $r=2 \; \mu$m. Note also that the initial part of the sectioned superposition will always agree with the single signal because during that time the two situations are identical.
\bef[th!]
\includegraphics[width=1\textwidth]{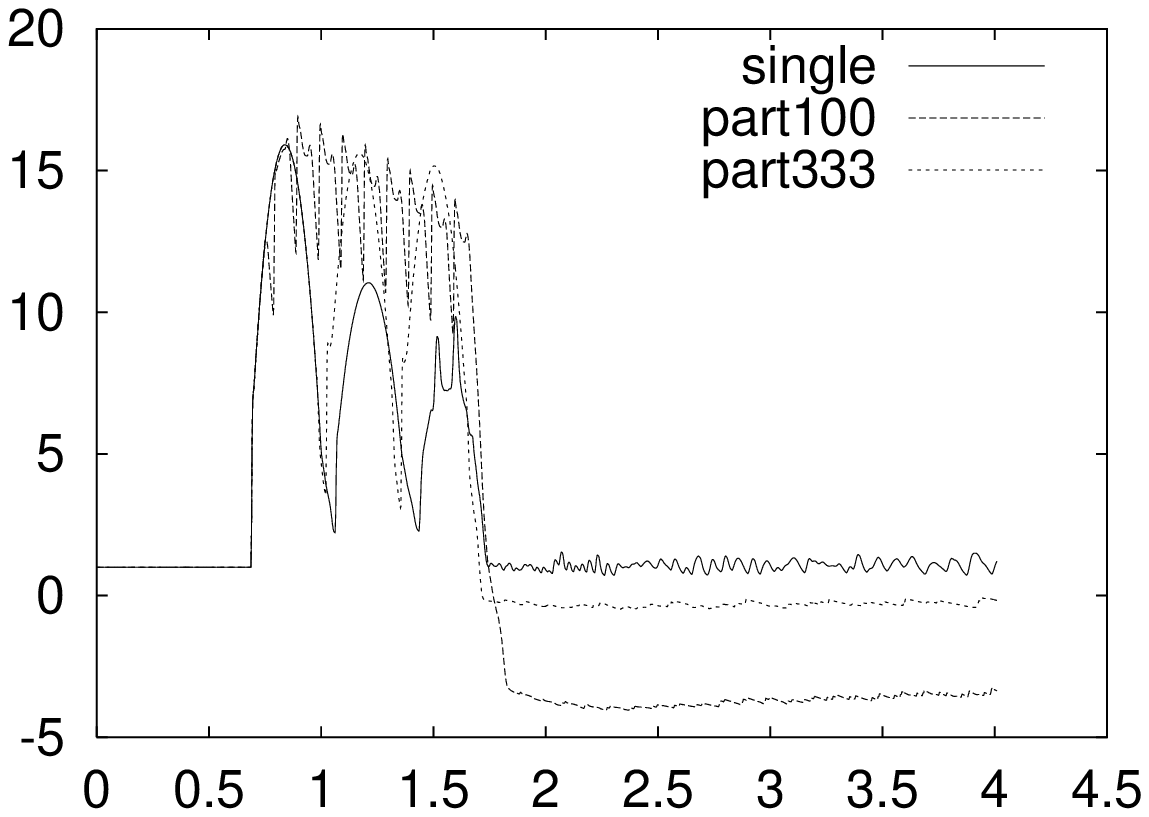} \\
\caption{Pressure at a location $u=2a$ as a function of time obtained from direct simulation and by applying the superposition principle with 10 sections of the original 1~ns laser pulse. The fluence of the laser pulse is $I_0 = 10^{-1}$~J/cm$^2$. In this case clear deviations between the two curves are are observed and the superposition principal is violated.}
\label{sp.j1}
\eef

To summarize, we have studied a general dynamic characteristic of linear systems: superposition. We found that for a melanosome irradiated by a laser with low laser fluence, the response behaves linearly and obeys superposition. That is, the total response of the system can be calculated by summing up individual responses from the system. As the fluence is raised, the system becomes increasingly more nonlinear, and the superposition principal is increasingly violated. We find a critical region for this transition for this system at a laser fluence of $I_0 = 10^{-3}$~J/cm$^2$, for this fluence deviations between the superimposed and the real signal are starting to deviate from one another. This work can lead to better understanding and quantifying the transition from linear to non-linear regimes in complex systems.

\begin{acknowledgments}
The author would like to thank the Air Office of Scientific Research (AFOSR) for funding through Grant No. F49620-03-1-0221.
\end{acknowledgments}

\bibliographystyle{unsrt}
\bibliography{supos}


\end{document}